\documentclass[preprint]{aastex}

\usepackage{graphicx}
\usepackage{amssymb}
\usepackage{amsmath}
\usepackage{xcolor}
\usepackage{bm}
\usepackage{xspace}
\usepackage{pgffor}
\usepackage{upgreek}
\usepackage{nicefrac}
\usepackage[normalem]{ulem}
\usepackage[single=true]{acro}

\newcommand{\be}{\begin{equation}} 
\newcommand{\ee}{\end{equation}} 
\newcommand{\nn}{\mbox{} \nonumber \\ \mbox{} }
\newcommand{\ba}{\begin{eqnarray}}
\newcommand{\ea}{\end{eqnarray}}
\newcommand{\om}{\omega}

\newcommand\eg{\textit{e.g.,\ }}

\newcommand{\Bf}{{magnetic field}}
\newcommand{\Bfs}{{magnetic fields}}

\newcommand{\NS}{neutron star}
\newcommand{\NSs}{{neutron stars}}

\newcommand{\ms}{magnetosphere}
\newcommand{\mss}{magnetospheres}

\newcommand{\LC}{light cylinder}
\newcommand{\Lf}{Lorentz factor}
\newcommand{\Lfs}{Lorentz factors}

\begin{document}

\title{Fast Radio Bursts from reconnection events in magnetar magnetospheres}

 \author{Maxim Lyutikov$^1$, Sergei Popov $^{2,3}$\\
$^{1}$ Department of Physics and Astronomy, Purdue University, 525 Northwestern Avenue, West Lafayette, IN, USA 47907\\
$^2$ Sternberg Astronomical Institute, 
Lomonosov Moscow State University, 119234, Russia\\
$^3$ Higher School of Economics, Department of Physics,
Moscow, 101000, Russia \\
}

 \begin{abstract}
 Lyutikov (2002) predicted ``radio emission from soft gamma-ray repeaters (SGRs) during their bursting activity". Detection of a  Mega-Jansky radio burst in  temporal coincidence with high energy bursts
 from a Galactic magnetar SGR 1935+2154 confirms that prediction.  Similarity of this radio event with Fast Radio Bursts (FRBs) suggests  that  FRBs are produced within magnetar \mss. We demonstrate that SGR 1935+2154 satisfies  the  previously derived constraints on the physical parameters at the  FRBs' {\it loci}.
 % (the \Bf\ and the  expected particle energy density). 
 Coherent radio emission is generated in the inner parts of the \ms\  at $r< 100 R_{\rm NS}$.  
 The radio emission is produced by the yet  unidentified   plasma  emission process, occurring      during  the initial stages of reconnection events.
\end{abstract}

%\maketitle

\section{Introduction}

Detection of Mega-Jansky radio burst from a galactic magnetar SGR 1935+2154 by CHIME and STARE2  (ATel 13681, 13681),  in  temporal coincidence with high energy bursts  observed by
 Integral (GCN 27666, ATel 13685), AGILE (ATel 13686), Konus-Wind (GCN 27669, ATel 13688), and
Insight-HXMT (ATel 13687)
 %Konus-Wind, Integral and AGILE  (ATel  13693, 13696, 13685, 13686)   
 identifies  the \ms\ of ``classical'' magnetar as the site of this event. Many features of this radio burst resemble Fast Radio Bursts (FRBs), suggesting  \mss\ of magnetars as FRB production sites
 (for review of FRBs phenomenology see
\cite{2018PhyU...61..965P, 2019PhR...821....1P, 2019A&ARv..27....4P, 2019ARA&A..57..417C}, and the catalogue in \cite{2016PASA...33...45P} \footnote{URL: http://frbcat.org/}; for  reviews on magnetars see \cite{2015RPPh...78k6901T,  2018ASSL..457...57G} and the catalogue in \cite{2014ApJS..212....6O} \footnote{URL: http://www.physics.mcgill.ca/~pulsar/magnetar/main.html}.

 The observations  are consistent with the concept that radio and X-ray bursts are generated during reconnection events in the magnetars \mss,  as suggested by  \cite{2002ApJ...580L..65L}.
 Conceptually,  magnetar  radio/X-ray flares  are  similar to Solar flares, initiated by the \Bf\ instabilities in the magnetars' \mss.
 (Radio emission from magnetars was further discussed by   \cite{eichlerradiomagnetar,2003MNRAS.346..540L,2006MNRAS.367.1594L,2015MNRAS.447.1407L}; 
   \cite{2010ARA&A..48..241B} discuss the electromagnetic signatures of  Solar flares.)

  \cite{2002ApJ...580L..65L}  suggested two spectral properties of magnetars' radio emission, one that was  confirmed and another relevant for the current problem. First, the prediction was  that magnetars will be brighter at higher radio frequencies since in twisted magnetospheres \citep{tlk}  the density and the plasma frequency can be much larger than in rotationally-powered  pulsars \citep{GoldreichJulian}. This was later confirmed: at tens of GHz magnetars are the brightest, though variable, pulsars \citep{2006Natur.442..892C,2014ApJS..212....6O,2017ARA&A..55..261K}. 
   Second,  a  downward frequency drift was predicted, by analogy with  Solar type-III radio bursts. This was not (yet?)  seen in magnetars, but   high-to-low frequency  drifting features  were indeed observed in the spectra of  the repeating FRBs \citep{2019ApJ...876L..23H}. The drifts  are  then the FRBs' analogues  of   Solar type-III radio burst \citep[but not in a sense of a particular emission mechanism,][]{2020ApJ...889..135L}. Related  concept is the  radius-to-frequency mapping in radio pulsars \citep{1977puls.book.....M,1992ApJ...385..282P}.
        
 The plasma parameters  required to produce FRBs are extreme, even when compared with brightest pulses from rotationally-powered pulsars \citep{2019arXiv190103260L}. \cite{2019arXiv190103260L} argued that 
 the requirements on plasma properties can be generally satisfied in magnetars' \mss. 
 In this paper we apply the more general conditions discussed by \cite{2019arXiv190103260L} to the particular case of radio bursts from  SGR 1935+2154.

 %\footnote{
To be clear, the term ``magnetar'' is used in two astrophysically separate settings: (i) powerful X-ray emitters, SGRs and AXPs, we term them ``classical magnetars'' \citep[eg][]{TD93,TD95,tlk02,2017ARA&A..55..261K} ; (ii) fast pulsar with high \Bf\ and high wind power, we term them ``millisecond magnetars''  \citep{Usov92,2006NJPh....8..119L,2008MNRAS.385.1455M}. In the former case the radiative energy comes from the energy of the \Bf, while  in the latter case from  its rotational energy (the strong \Bf\ just serves to quickly transform the rotational energy into the wind). 
% Many models of FRBs employ the second scenario, using the word ``magnetar'' to mean  central source producing  rotationally-powered winds \citep{2019ApJ...886..110M,2019MNRAS.485.4091M}. 
FRBs are not  rotationally powered   ``fast-rotation pulsar with high \Bf''-magnetar concept, but are magnetically-powered  ``classical'' magnetars sources.
We also note that long term periodicity observed in some FBRs \citep{2020arXiv200110275T,2020arXiv200303596R} is consistent with {\it mildly powerful} \NS, not a millisecond magnetar-type \citep[][wrote ``The observations are consistent with magnetically powered radio emission originating in the magnetospheres of young, strongly magnetized neutron stars, the classical magnetars.'']{2020ApJ...893L..39L}.

%In the case of FRB from SGR 1935+2154,  the temporal  coincidence of radio and high energy emission implies spatially coincident {\it loci}. This is inconsistent with the production of FRBs in magnetars'   winds \citep[as suggest, \eg\ by][]{2014MNRAS.442L...9L,2017ApJ...843L..26B}.

%}.
 
\section{Observational time-line and  best contemporaneous  interpretation of FRBs}

As the origin of FRBs is being settled down, let us have a quick look at  the observational time line and the best contemporaneous  interpretations, as viewed now:
\begin{itemize}
\item ``Early years'': from \cite{2007Sci...318..777L}  to \cite{2013Sci...341...53T}.
Time of many hypothesis, no leaders. Among others, merging {\NS}s  are discussed   \citep[see early discussion of transients from NS-NS mergers by][]{2001MNRAS.322..695H,2011PhRvD..83l4035L}, and later proposals particularly for FRBs by \citet{2013PASJ...65L..12T,2014A&A...562A.137F}. The main argument  for NS-NS merges was that the active stage lasts $\sim R_{NS}^{3/2}/  ( G M_{NS} )^{1/2} \sim$ milliseconds, matching FRB duration. Magnetar origin (as well as scaling of Crab giant pulses with rotational energy losses) is proposed, but mostly ignored \citep{20077arXiv0710.2006P}.
%\item ``Early years'': from \cite{2007Sci...318..777L}  to \cite{2013Sci...341...53T}.  Merging {\NS}s are best guess \citep{2001MNRAS.322..695H,2011PhRvD..83l4035L,2013PASJ...65L..12T,2014A&A...562A.137F} (main argument is that the active stage of NS-NS merger lasts $\sim R_{NS}^{3/2}/  ( G M_{NS} )^{1/2} \sim$ milliseconds). 
%\item ``Early years'': from \cite{2007Sci...318..777L}  to \cite{2013Sci...341...53T}.
% Time of many hypothesis including exotics (cosmis strings, evaporating black holes).
%Magnetar origin (as well as scaling of Crab giant pulses with rotational energy losses) is proposed, but mostly ignored \cite{2013arXiv1307.4924P}.
\item ``High rates years'': from \cite{2013Sci...341...53T} to \cite{2016Natur.531..202S}. Rates, $\sim 10^4$ per sky per day, are well in excess of NS-NS merger rates;  hence FRBs most likely involve non-destructive events.  Burst energetics requires NS-type \Bfs\ \citep{2016MNRAS.462..941L}. Magnetically powered magnetar flares \citep{2013arXiv1307.4924P} or  rotationally powered giant pulses \citep{2016MNRAS.462..941L,2016MNRAS.458L..19C,2016MNRAS.457..232C} are best guesses. 
\item ''The Repeater years'':    from  \cite{2016Natur.531..202S} to Apr 28 2020.   FRBs must come from non-destructive  sources. Rotationally powered giant pulses are excluded by energetics \citep{2017ApJ...838L..13L}. Magnetars are best guess, but there are mild  observational constraints \citep{2019arXiv190103260L}. Observations of  consistent spectral drifts \citep{2019ApJ...885L..24C,2019ApJ...876L..23H} point to \mss\ of  NSs  as {\it loci} of emission \citep{2002ApJ...580L..65L,2020ApJ...889..135L}.
\item CHIME years \citep{2018ATel11901....1B}: 2018- 
\end {itemize}

 \section{SGR 1935+2154 and constraints on local  FRB properties}
 \label{constraints}
 
 \subsection{Properties of SGR 1935+2154}
 
SGR 1935+2154 was discovered due to a weak $\sim 0.3$~s burst in 2014 by Swift \citep{2014GCN.16520....1S}. Further analysis demonstrated that the source might be a magnetar \citep{2014GCN.16522....1L}.
Pulsations with the period 3.245 s were recorded the same year by Chandra \citep{2014ATel.6370....1I}.

Dedicated Chandra and XMM-Newton observations, as well as analysis of archival data, allowed to measure precisely parameters of the magnetar \cite{2016MNRAS.457.3448I}. X-ray luminosity in quiescent state is $\sim$few$\times 10^{34}$~erg~s$^{-1}$, i.e. slightly above  spindown luminosity $\dot E_\mathrm{rot} =1.7\times10^{34}$~erg~s$^{-1}$. Spectrum can be fitted by a combination of the blackbody and power-law radiation or double black body \citep{2016MNRAS.460.2008K,2017ApJ...847...85Y}. The source was also identified in pre-burst archive data of XMM-Newton with luminosity $\geq 10^{34}$~erg~s$^{-1}$, i.e. with $L_\mathrm{x}\approx \dot E_\mathrm{rot}$, which put the source on the boundary between radio-silent and radio-loud magnetars (the former typically have $L_\mathrm{x} > \dot E_\mathrm{rot}$. Early attempts to detect radio emission from SGR 1935+2154 after 2014 outburst resulted just in upper limits $\sim0.1$~mJy.  

SGR 1935+2154 is associated with a shell-type supernova remnant G57.2+0.8. This allows to determine the distance of $12.\pm 1.5$~kpc   \citep{2018ApJ...852...54K}.\footnote{We use this value in the paper, however,  recently  \cite{2020arXiv200503517Z}  estimate the distance as $6.6 \pm 0.7$~kpc, i.e. twice smaller.} In addition, an age estimate $\sim$few tens of thousand years (with the more probable value 41 000 yrs) was obtained. These number a roughly compatible with the spin-down age $\sim3.6$~kyr. (It is tempting to speculate that the age determined via SNR studies is slightly larger than the spin-down age due to magnetic field emergence, see \cite{2013ApJ...770..106B} and references therein.)
A pulsar wind nebula is suspected due to diffuse emission detected in X-rays \cite{2016MNRAS.457.3448I}.

With a period $P=3.24$ seconds, and period derivative $\dot{P} = 1.4 \times 10^{-11}$ \citep{2014ApJS..212....6O} the surface \Bf\  $B_{NS} $ and the fields at the light cylinder $B_{LC}$ evaluate to
\ba && 
 B_{NS} = 2.2 \times 10^{14} {\rm G}
 \nn &&
 B_{LC} = 60 {\rm G}
 \label{1}
\ea

  \subsection{Radio burst from  SGR 1935+2154: a weak FRB}
  
  Based on the data by STARE2 (fluence $\sim$MJy ms) and CHIME (duration about 40 ms, due to two 5 ms bursts and 30 ms interval between them) we obtain:
\be
L_{\rm{R}} =0.66\times 10^{37} \left(\frac{d}{12.5\,\rm{ kpc}}\right)^2 \left(\frac{\nu}{1.4 \,\rm{GHz}}\right) \left(\frac{\rm{fluence}}{1\, \rm{MJy}\times\rm{msec}}\right)\left(\frac{\rm{duration}}{40\,\rm{msec}}\right)^{-1}\,\rm{erg}\,\rm{s}^{-1}.
\label{LR}
\ee
This estimate of intrinsic luminosity depends sensitively on the uncertain location of the source within the radio beam! Intrinsic luminosity could be higher!

Taken at face value, luminosity  (\ref{LR})   is  lower than typical intrinsic radio luminosities of FRBs, which are estimated at $\sim 10^{39}$~--~$10^{42}$~erg~s$^{-1}$ for isotropic emission. (If one uses the value of fluence reported by CHIME --- few kJy$\times$ms, ---  then the radio luminosity goes down by $\sim3$ orders of magnitude.)

 %For comparison, the nearest  FRB 180916.J0158+65 at 149 Mpc produces Jansky-level bursts \cite{2020Natur.577..190M}. If SGR 1935+2154 were at the distance of FRB 180916.J0158+65 it would have produces FRB at $\sim 10$ milli-Jansky level, about two orders of magnitude weaker.
  Thus, the  intrinsic luminosity of FRB from  SGR 1935+2154, Eq. (\ref{LR}), as well as the  brightness temperature $T_b$,   Eq. (\ref{Tb}), are still short of the cosmological FRBs, that require \eg\ peak luminosity above $10^{40}$  erg s$^{-1}$. 
   Given the spread of FRBs' intrinsic luminosities, it is reasonable to assume that the radio burst from  SGR 1935+2154 represent a lower end of a broad distribution of FRBs' power:  ``only some special types of magnetars  can produce [cosmological] FRBs'' \citep[as argued by][]{2020ApJ...889..135L}.  \citep[We also note that previous observations did not detect radio bursts from SGR 1935+2154][this further indicates that there is a range of luminosities of radio bursts]{2017ApJ...847...85Y}. 
   %(We thank Vicky Kaspi and Jason Hessels for pointing these issues to us.) 
   (We thank  Jason Hessels for pointing these issues to us.)

%\cite{2020Natur.577..190M} also find a burst  from SGR 1935+2154 with higher fluence of 0.2 Jy ms (B2; see their Table 1;  also, this is close to their sensitivity limit). Fainter bursts are still being produced:\citep{2019ApJ...877L..19G}  did not see see a turnover in the fluence towards weak bursts. (We thank Jason Hessels for pointing  this.)

  Note, that recently numerous weak short radio bursts were detected from another galactic magnetar XTE J1810-197 \citep{2019ApJ...882L...9M}. However, no accompanying high energy activity was registered. These events had relatively low radio fluxes ($\lesssim 10$~Jy down to $\sim$mJy), so intrinsic luminosity is $\sim10^{11}$ times lower than, for example, in bright bursts of FRBs 121102. 

 \subsection{Emission radius of the radio burst}
 
  In this section we demonstrate that the parameters of SGR 1935+2154 are generally consistent with requirements to produce bright coherent FRB within the magnetar \ms\ \citep{2019arXiv190103260L}. 
  %For numerical estimates we normalize the distance to 9kpc, the observed FRB flux to 1MJy  and duration to one millisecond.

% With a period $P=3.24$ seconds, and period derivative $\dot{P} = 1.4 \times 10^{-11}$ \citep{??} the spin-down time 
% \be
% \tau_{sd} = \frac{P}{2 \dot{P}} = 3.8 \times 10^3 \,  {\rm yrs} 
% \ee
% matches ``classical magnetars''.
 
%  The spindown luminosity evaluates to
% \be
% L_{sd} = I_{NS} \Omega \dot{\Omega} =1.6 \times 10^{34} {\rm erg s}^{-1}
%\ee

 GCN  27668 reported that during the active phase the persistent X-ray luminosity is  $L_X \sim$ few $\times 10^{35}$ erg s$^{-1}$. This is  larger than the spindown luminosity, placing SGR 1935+2154 among ``classical magnetars''.
What is even  more exciting, the peak (isotropic equivalent) radio luminosity  to  (\ref{LR}) exceeds the spindown power. 
Most importantly,  this radio dominance excludes rotationally-powered emission as argued by  \cite{2017ApJ...838L..13L}.

Given the \Bfs\ (\ref{1}) 
the  cyclotron frequency  at the  \LC\ is 
\be
 \frac{\om_{B, LC}}{2 \pi} = 2 \times 10^8 {\rm Hz}
 \ee
 Thus, the emitted radio waves' frequency, if FRBs originate within the \ms, is typically below the cyclotron frequency. This is highly important for a consistent description of FRBs emission \citep{2019arXiv190103260L}. \citep[Also, the  cyclotron absorption, if any,  is likely to be in the magnetosphere, not the 
wind.][]{2001MNRAS.325..187L}

The radiation energy density at the source and the observed  brightness temperature evaluate to  \citep[we neglect  for simplicity possible relativistic motion][] {2019arXiv190103260L}
\ba &&
u_r = \frac{ \nu F_{\nu} d^2}{c^3 \tau^2}
\nn && 
T_b =\frac{u_r}{2\pi k_B} \lambda^3 = 8 \times 10^{30} \, {\rm  K} 
\label{Tb} 
\ea

The equipartition \Bf\ is
\be
B_{eq} = \sqrt{ 8 \pi u_r} =  \sqrt{8 \pi} \frac{ (\nu F_\nu)^{1/2} d}{c^{3/2} \tau}= 2 \times 10^6 {\rm G}
\ee
The equipartition \Bf\ is a lower estimate on the \Bf\ in the emission region \citep{2016MNRAS.462..941L}. Given the surface \Bf\  (\ref{1}), this limits the emission radius to
\be
\frac{r}{R_{NS}} \leq  \left( \frac{B_{NS}}{B_{eq} }\right)^{1/3} = 300
\label{rr}
\ee
Thus, emission is produced in the inner parts of the magnetar \ms.

Laser intensity parameter $a$ \citep{1975OISNP...1.....A} evaluates to
\be
a = \frac{ e F_\nu^{1/2} d}{\sqrt{\pi} m_e c^{5/2}  \sqrt{\nu} \tau} = 5 \times 10^3
\ee
As discussed by \cite{2017ApJ...838L..13L,2019arXiv190103260L,2020arXiv200109210L}, such large intensity parameter  requires large \Bf: 
otherwise  coherently emitting particles will have dominant ``normal'' losses (synchrotron and inverse Compton losses). In high \Bf\ instead of large oscillations with momentum $p_\perp \sim a m_e c$ coherently emitting particles experience mild  $E \times B$ drift. 
This requires $\om_B \geq \om$; this condition is satisfied at radii (\ref{rr}).

 \subsection{Plasma parameters at the radio burst production cite}
 
Magnetar \mss\ are non-potential configurations \citep{1994MNRAS.267..146L,tlk02},  twisted by the Hall drift in the NS's crust \citep{RG,RheinhardtGeppert02,Cummings,Reisenegger07,2013PhRvE..88e3103L,2014PhPl...21e2110W,2015MNRAS.453L..93G}. Current-carrying magnetospheric charges moving with \Lf\  $\Gamma$   
have a rest-frame density of 
\be
n' = \Delta \phi \frac{B}{2 \pi e r \gamma_s}
\ee
where $ \Delta \phi$ is s typical twist angle \citep{tlk02}, and $n'$ is plasma density in its center-of-momentum frame.

Radiative efficiency less than unity  (radiation energy density  less than plasma energy density) then requires \citep{2019arXiv190103260L}
\be
 \gamma \gamma_s^3 \geq \frac{ e \nu^3 r k_B T}{ m_e c^5 B \Delta \phi}= 6\times10^{13}   \nu_{GHz}^3  \frac{1}{\Delta \phi}  \left( \frac{r}{R_{LC}}\right)^{4}
 \label{tt}
 \ee
 where $\gamma$ is  the spread of the \Lfs\   in the center-of-momentum frame and $\gamma_s $ is the bulk streaming \Lf. The extremely high brightness temperatures in FRBs likely  involves relativistic plasmas, $\gamma, \, \gamma_s \gg 1$.  (\cite{2019arXiv190103260L} also included effects of radiation anisotropy; this decreases the demands on plasma energy content.)
 
Near the \NS\ surface the condition (\ref{tt})  evaluates to 
\be
 \gamma \gamma_s^3 \geq 10^{-3}   \nu_{GHz}^3  \frac{1}{\Delta \phi}  
 \ee
 This can be easily  satisfied even for very mild twist $\Delta \phi \leq 1$.

We conclude that the inner part of the   \ms\ of SGR 1935+2154, the \Bf\ and expected plasma density,  satisfies the physical requirements to produce high brightness coherent emission.

 \section{Solar physics of magnetars }

\cite{2006APS..APR.X3003L} discussed similarities between Solar flares and magnetar phenomenology,   the  ``Solar physics of magnetars''. Detection  of radio bursts contemporaneous with high energy bursts further strengthens this analogy.

   Light curves and radio-X-ray  relative timing may also hold clues to interpretation of flares.
   The first radio pulse is delayed by ~100-200 ms relative to the X-ray
emission onset in the Konus-Wind softest energy band (GCN 27669, ATel
13688).
   %Konus-Wind data (ATel 13693, 13696) indicate that  in the harder 80-230 keV band there is not delay, but in the softer band X-rays star $\sim 150-200$ ms earlier.
   This behavior is qualitatively similar to the solar flares, where soft X-ray emission  starts to rise before the prompt flare \cite[\eg\ Fig. 2 in][]{2008LRSP....5....1B}. In addition, in Solar flares soft X-rays continue longer than the prompt non-thermal X-rays and microwaves spikes  \citep[``Neupert effect''][]{1968ApJ...153L..59N}.
   Indicatively, magnetar giant flares show similar hard-to-soft evolution \citep{palmer,2017ARA&A..55..261K}.
     In case of the Sun the interpretation is that  the soft X-rays  originate  from  plasma heated by the primary  flare's   electrons.
   In the case of magnetars we expect that reconnection events lead to abundant pair production, and ``pollution'' of the acceleration region, as we discuss next.

According to   GCN  27668,  Integral detected a peak flux of
$\sim 10^{-6}$  erg cm $^{-2}$ s$^{-1}$  (which can be slightly underestimated), this corresponds to peak luminosity $L_\gamma = 9\times 10^{39}$ erg $^{-1}$. For a duration of  $\tau _\gamma \sim 10$ msec the compactness parameter evaluates to
\be
l_c =\frac{\sigma_T L_\gamma }{m_e c^4 \tau _\gamma}\approx 10^3
\label{lc}
\ee
Thus, we expect  abundant pair production following the gamma-ray flare. 

  We hypothesize that the radio emission is generated during the initial stage of magnetospheric reconnection, while the \ms\ is still relatively clean of the pair loading. The  giant $\gamma$-ray flare from the magnetar SGR 1806 - 20 had a rise time of only $200$ micro-seconds \citep{palmer}, matching the duration of the radio flare.
  In a possibly related study  of relativistic reconnection by \cite {2017JPlPh..83f6301L, 2017JPlPh..83f6302L,2018JPlPh..84b6301L}, it was found that in highly  magnetized plasma the reconnection process driven by large scale stresses (magnetically-driven collapse of an X-point) has an initial stage of extremely fast acceleration, yet low level of magnetic energy dissipation. This initial stage of reconnection may produce unstable particle distribution in the yet clean surrounding, not polluted by pair production.
  
   In the  case of mixed magnetar/radio pulsar PSR J1119-6127 X-ray bursts actually lead to suppression of radio emission \citep{2017ApJ...849L..20A}. We interpret this as follows:  (i)   rotationally-powered radio emission in  PSR J1119-6127 was polluted by pair production during flares, and shut off; (ii) in  SGR 1935 we are dealing with a different radio emission mechanism, reconnection-driven.  Neither are understood, but the suggestion is that at early times during reconnection the plasma is still clean to allow unstable particle distribution to be created. Later in the flare it will also be suppressed. Thus, one of the predictions of our model  is that radio precedes high energy, approximately  by a fraction of burst duration.

   We favor the ``Solar flare'' paradigm of magnetar flares  driven by {\it plastic deformations} of the crust  \citep{2012MNRAS.427.1574L,2015MNRAS.447.1407L}, as opposed to ``starquakes" model of   \citep{TD93}.  \cite{2012MNRAS.427.1574L} demonstrated that magnetically-induced cracking is not possible:  burst and flares are more naturally produced as magnetospheric events, analogous to Solar flares \citep{2015MNRAS.447.1407L}.
  
Importantly,  the total energy budget of the burst, $\sim 10^{39}$ erg, can be easily accommodated. For surface \Bf\ (\ref{1}) the required volume of dissipated magnetic  energy corresponds to only few tens of meters cubed, much smaller than the size of the \NS. The magnetic energy budget in magnetars could in fact  be higher than the one estimated from the surface fields due to the large required toroidal fields in the crust
  \cite[eg][]{FlowersRuderman,2006A&A...450.1077B,2013MNRAS.434.2480G}.
  
\begin{figure}[h!]
\centering
\includegraphics[width=.99\textwidth]{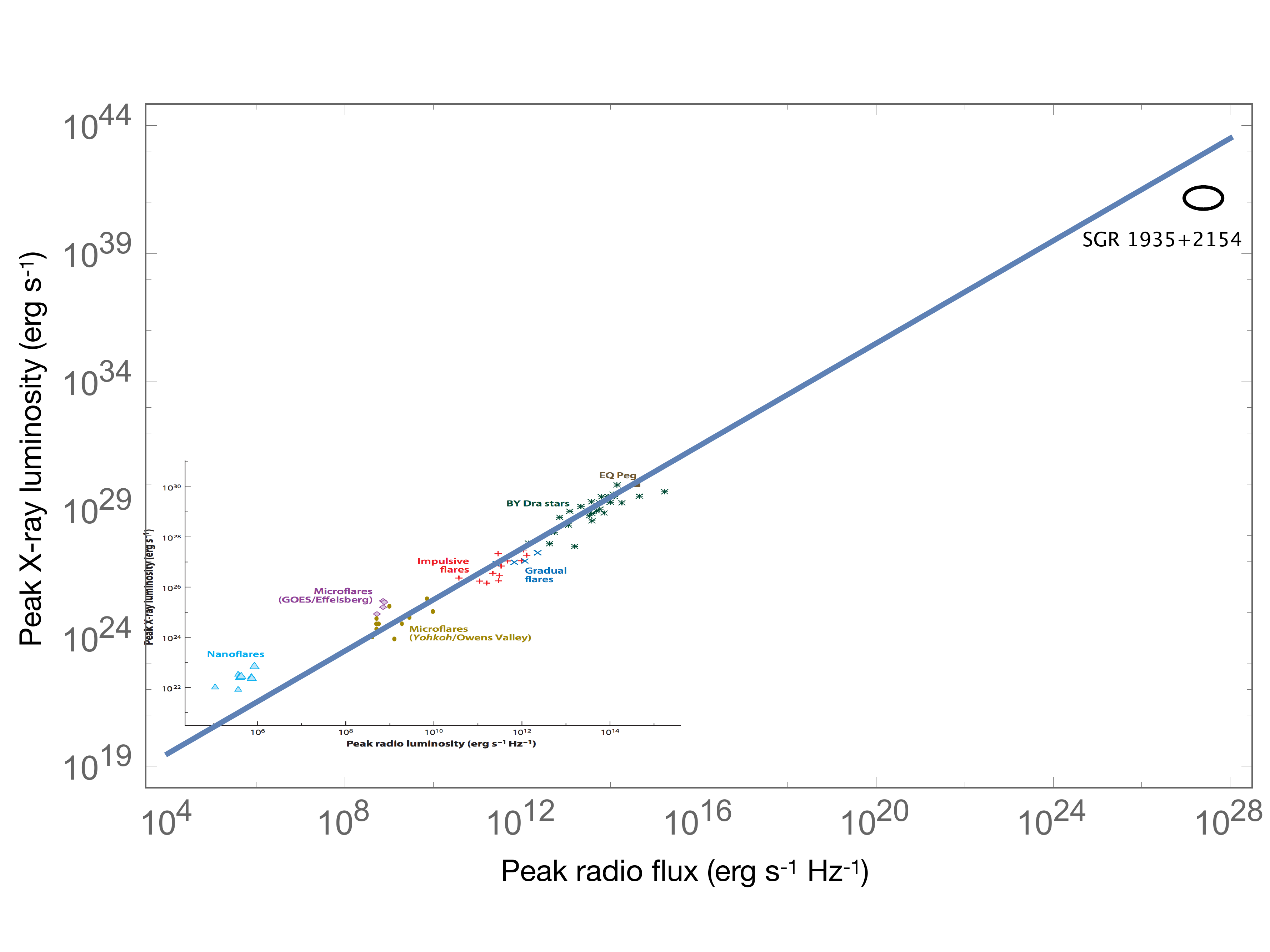}
\caption{X-ray and centimeter radio luminosities of stellar flares \protect\citep{2010ARA&A..48..241B} and radio (STARE2, ATel 13681) and X-ray (Konus-Wind, ATel 13688) flares from SGR 1935+2154. 
(Figure to be updated with original data once they  become available after the end of the lockdown).
}
\label{WD-RD-int} 
\end{figure}
We  also note that the high energy and radio burst from SGR 1935+2154 lies not far from the so-called 
 G\"{u}del-Benz relationship, which relates the thermalized X-ray luminosity generated by magnetic reconnection in stellar flares to the nonthermal, incoherent, gyrosynchrotron radio emission that results from particle acceleration  \citet[for a review][also see a comment after Eq. (\ref{LR}) about the uncertainty in radio luminosity of SGR 1935+2154]{2010ARA&A..48..241B}. This relationship is given by
$ {L_{X} / \tilde{L}_{R}}\sim 10^{15.5\pm 0.5}~[Hz], $
where $L_{X}$ is the X-ray luminosity and $\tilde{L}_{R}$ is the radio flux per unit frequency. Using peak X-ray flux (ATel 13688)
of $F_X = (9\pm 2)\times 10^{-6}  $erg~cm$^{-2}$~$s^{-1}$ (in the 20 - 200 keV energy range)
 the ratio estimates to $6 \times 10^{13}$: not too far off!
 
G\"{u}del-Benz correlation is usually interpreted that  first electrons are accelerated to non-thermal velocities and emit radio, then these particles are thermalized and emit thermal X-rays. The magnetar SGR 1935+2154 adds another point, with an important caveat that the radio emission in this case is coherent. (Interestingly, it also lies off to the ``expected'' side from the lower energetics fit: the radio is too bright, as expected for coherent emission.) 
Though the microphysics of this relation is far from clear, is it at least consistent with the concept that accelerating mechanism puts first energy into nonthermal particles that produce radio, and then that energy is thermalized, producing X-rays.

Solar X-ray flares show a power law distribution of energies, extending over many orders of magnitude \citep{2010ARA&A..48..241B}. The value of the power law index and  the origin of this distribution is not clear: a self-organized criticality  (sand pile model) is one possible explanation. Qualitatively, magnetic configurations are  nonlinear systems that show a ``threshold''-type behavior: slow evolution below threshold gives rise to exponential growth after the instability threshold is crossed.
In application to FRBs, we similarly  expect a broad range of bursts' energetics.

Another important point is the similarity of the
polarization properties of the magnetar  burst (ATel 13699)  and of  original repeaters FRB 121102 and FRB 180916.J0158+65
\citep{2018Natur.553..182M,2019ApJ...885L..24C}. They all show $\sim 100\%$, linear polarization,  negligible  circular polarization and a consistently flat PA.  
This consistent polarization patterns, as well as consistently  drifting  spectral features imply a kind of ``confining structure":   the magnetospheric \Bfs. These consistent features also imply that the duration of the pulses are  intrinsic, as opposed to being due to a beam that is longer lived and sweeps past the line of sight. (We thank Jason Hessels for stressing this similarity.)

 The origin of the coherent emission of FRBs remains a mystery (as well as that of regular pulsars). 
 As discussed by   \cite{2016MNRAS.462..941L} we can
identify {\it three} types/mechanisms of radio emission in \NSs: (i) normal
pulses, exemplified by Crab precursor, \cite{1996ApJ...468..779M}; (ii) giant pulses, exemplified by Crab Main
Pulse and Interpulse \citep{1995ApJ...453..433L,popov06,2019MNRAS.483.1731L,2019ApJ...876L...6P}; (iii) radio emission from magnetars
(coming from the region of close field lines, \eg\ \cite{2002ApJ...580L..65L,eichlerradiomagnetar,2006Natur.442..892C}. FRBs should be of type-iii radio emission.

The lack of understanding of
mechanisms of radio emission from  normal pulsar is a major impediment to the future 
progress 
\citep[\eg][]{1995JApA...16..137M,1999MNRAS.305..338L,1999ApJ...521..351M,2015SSRv..191..207B}. 
Note, that coherent curvature emission by bunches,   popular in the early years of pulsar research \citep{1971ApJ...170..463G,1977ApJ...212..800C}, is not considered a viable
emission mechanism \citep{1977MNRAS.179..189B,1990MNRAS.247..529A,1992RSPTA.341..105M,1999ApJ...521..351M}. In addition, in case of repeating  FRBs the absence of periodicity excludes narrow region of open field lines for the production of coherent emission. In a magnetar paradigm  reconnection events  occurring in broad regions of the \ms\ may hide the rotational period.

 \section{Discussion and expectation}
 
  Detection of MJy radio burst from a Galactic magnetar and many similarities to FRBs points to magnetically powered ``classical magnetars'' 
%  \footnote{In a  \cite{TD93} sense, not  rotationally powered ``millisecond magnetars'', \cite{Usov92}}
   as the {\it  loci} of FRBs.  Temporal coincidence between radio and high energy emission, down to milliseconds, further limits the FRB {\it loci} to the \mss\ of magnetars (as opposed to winds, see below).

   Radio emission from reconnection events in magnetars' \mss\  was previously predicted/discussed by \cite{2002ApJ...580L..65L,2003MNRAS.346..540L,2006MNRAS.367.1594L,2015MNRAS.447.1407L}.
In particular, drawing on analogies with solar flares, \cite{2002ApJ...580L..65L}  predicted that coherent radio emission  may be emitted in SGRs during X-ray bursts. Thus, these event are reconnection-driven emission processes occurring in magnetars' \mss. In addition, \cite{2002ApJ...580L..65L} argued that emission should have downward  drifting central frequency in analogy with solar type III radio bursts. Though such drifts have not yet been seen in magnetars, they were indeed detected in FRBs \citep{2019ApJ...876L..23H,2019Natur.566..235C,2019arXiv190803507T,2019arXiv190611305J,2020ApJ...889..135L}. Observations of such drifts in magnetar bursts would  further strengthen  FRB-magnetar connection.
      
  In this paper we demonstrate that extremely demanding conditions on plasma parameters at the sources of FRBs, discussed previously by \cite{2019arXiv190103260L}, can be {\it easily} accommodated in the case of SGR 1935+2154, with no extreme assumption about the expected local plasma parameters. (This is due to the identification of the source, giving us estimates of distance, period and  period derivative.) On the one hand, a particular radio burst from SGR 1935+2154 is at least $\sim 100$ times less powerful that the weakest FRB detected. On the other hand, magnetar SGR 1935+2154 is not a particularly special magnetar in any respect. There is a ``room'' in parameter space to produce brighter radio bursts.
  
  Mildly  optically thick plasma, Eq. (\ref{lc}),  is also expected to produce thermal  high energy spectrum. Data from  Konus-Wind are inconsistent with power law  (ATel 13693, 13696, GCN 27669). Also, double black-body detected by Konus-Wind (GCN 27669) resembles double black-body spectrum detected previously by 
\cite{2016MNRAS.460.2008K,2017ApJ...847...85Y}. This is likely due to  polarization-dependent  emission transfer in magnetically-dominated  (Lyutikov, in prep.)

  Perhaps the first FRB-magnetar connection was discussed by   \cite{20077arXiv0710.2006P} who suggested that FRBs can be due to hyperflares on magnetars, whose  X/$\gamma$-rays emission is  undetectable  from distances $\geq 100$~Mpc. It was shown that from the point of view of total energy budget, rate, timescale, and  lack of counterparts properties of energetic flares of extragalactic magnetars are consistent with the hypothesis that they produce millisecond radio flares. The mechanism of radio emission production was not specified, but the authors used the model by \cite{2002ApJ...580L..65L} to obtain basic numbers and to speculate about radio flares of different energy related to correspondingly different X/$\gamma$-ray bursts. When the paper by \cite{2013Sci...341...53T} appeared, \cite{2013arXiv1307.4924P} noticed that  the magnetar hypothesis fits well new data, too. 
  %Since that time this suggestion was considered among the most relevant idea of the the FRB origin. May be now it can verified with observations of weak bursts of the Galactic magnetars. 

In terms of astronomical locations, the picture is not clear: the two well-localized repeaters (FRB 121102 and FRB 180916.J0158+65) are both found coincident with star-forming regions 
\citep{2017ApJ...843L...8B,2020Natur.577..190M}. This is consistent with magnetar origin.  At the same time, the (apparent) non-repeaters are not obviously associated with star formation \citep{2019Sci...365..565B}.  One possibility is that  the non repeaters are much older NSs that only very occasionally produce a bright burst (we thank Jason Hessels for pointing this to us).

Magnetospheric origin  of  radio emission can hardly be questioned now. Magnetar giant flares \citep{palmer} are clearly magnetospheric events, as indicted by the periodic tail oscillations with the previously known spin period.  Lower energy X-rays bursts are just weaker events, corresponding, approximately, to the initial spike of the giant flare \citep{2017ARA&A..55..261K}. Temporal coincidence between radio and high energy, down to few milliseconds, implies then co-spacial origin.

   Association of an FRB  simultaneous with a relatively weak X-ray bursts of only $\sim 10^{39}$ ergs,  from a ``nothing special'' magnetar,  is inconsistent with the wind model of \cite{2014MNRAS.442L...9L,2017ApJ...843L..26B}, see though  \cite{2020arXiv200505283M}.  Qualitatively, in those models emission comes from large distances, requiring large post-shock \Lfs,  as large  as $\Gamma \geq 10^4$, to produce short duration.  The pre-shock \Lf\ should be even higher. Thus, the initial required total energy at the source, at the time of shells' launching, is by a factor $\Gamma$ --  orders of magnitude --  larger than the observed X-ray flare.  Such energy releases (to produce multiple peaks numerous shells are needed)  would require a disruption of a large part of the \ms; this was  not observed in the case of the Galactic magnetar SGR 1935+2154. (In comparison, the present model needs just about a football field of magnetic energy.)

%   In that  model, emission comes from the distance of few $\times 10^{16}$ cm: to produce millisecond duration the {\it post-shock} \Lf\ should be of the order of $\Gamma \sim 10^4$. The pre-shock \Lf\ should be even higher, $\Gamma \gg 10^4$. 
%    In the  concept of colliding shells in Gamma Ray Bursts, if two shells of \Lfs\ $\Gamma_1$ and $\Gamma_2\gg \Gamma_1$  collide, the  relative energy of the relative motion is only  $\propto \Gamma_2/(2 \Gamma_1^2)$. 
%Thus, the initial required total energy at the source  becomes of the order of the giant flare energy $\geq 10^{46}$ erg. Such energy releases (to produce multiple peaks numerous shells are needed)  require a disruption of a large part of the \ms; this was  not observed in the case of the Galactic magnetar SGR 1935+2154. (In comparison, the present model needs just about a football field of magnetic energy.)    Also, the required \Lfs\  $\Gamma \gg 10^4$ cannot be achieved even in the cleanest fireballs   due to remaining loading by pairs \cite[\eg][]{2004RvMP...76.1143P}.    In addition,  the observed spectral peak  of $\sim 10-30 $ keV (ATel 13688, GCN 27669) would then correspond to $\sim 0.1$ eV photons in the plasma rest frame: such low energy photons will be emitted by electrons in slow cooling regime, further boosting the overall energy requirement (since the energy of accelerated particles will  be mostly lost to adiabatic expansion, not radiation), and contradicting short duration of a burst.   
  
Looking forward,   we expect that  more pulsar-like   phenomenology to be discovered in FRBs. Though   the  energy sources in pulsars and magnetars are  different  (rotational energy versus the magnetic field), the overall dominating \Bf\ is expected to  impose many similar observational  effects. 
The most obvious is the periodicity, reflecting the rotational period of the \NS. Another 
 prediction is the polarization swings through the pulse \citep[rotating vector model is a corner stone of pulsar phenomenology][]{1969ApL.....3..225R}.
 Polarization swings are expected in case of emission originating on magnetars' close field lines, but  shorter  rotational period NSs produce large PA swings \citep{2020ApJ...889..135L}. PA swings were not  seen in this particular case (ATel 13699), presumably due to the  fairly long period of a \NS; still a high degree of linear polarization  is consistent with highly structured magnetars' \ms.
Curiously, microseconds-long giant pulses from Crab pulsar, with approximately similar relative pulse duration, do sometimes show flat polarization angle. Giant pulse are also likely to be generated  in reconnection events, though outside the light cylinder \citep{2016ApJ...833...47H,2016MNRAS.463L..89C,2017SSRv..207..111C,2019ApJ...876L...6P}. 
We also expect detection  of narrow spectral features and frequency drifts in magnetar radio bursts, akin to the ones seen in FRBs. This   will further solidify the association.
Finally, we expect that radio emission precedes the high energy, by a fraction of burst duration, few milliseconds.

Finally, the identification of FRBs with magnetars implies that we  are not likely to detect cosmological FRBs by all-sky X-ray/gamma-ray monitors. As discussed by \cite{2016ApJ...824L..18L}, given the  magnetars' X-ray flares maximal radio power of $\sim  10^{47}$  erg sec$^{-1}$ \citep{palmer}, they can be detected   only to $\sim 100$ Mpc. On other other hand,  sensitivity of imaging high energy telescopes may allow observations of contemporaneous FRB/gamma-ray flares  in the previously identified   repeaters out to $\sim$ Gpc
\citep[see also][]{2017ApJ...846...80S,2020arXiv200406082S,2019ApJ...879...40C}. Also, simultaneous detection in optical may be possible \citep[\eg\ by ''shadowing'' of CHIME field by an optical telescope,][]{2016ApJ...824L..18L}.

\section*{Acknowledgements}

We would like to thank members of  the Konus-Wind team  (Dmitry Svinkin, Anna Ridnaia, Dmitry Fredereriks) and CHIME project (Vicky Kaspi and Shriharsh Tendulkar). We also thank Arnold Benz, Roger Blandford, Jason Hessels, Amir Levinson and Alexander Philippov for discussions.

ML would like to acknowledge support by  NASA grant 80NSSC17K0757 and NSF grants 10001562 and 10001521. 
SP acknowledge the support from the Program of development of M.V. Lomonosov Moscow State University (Leading Scientific School ``Physics of stars, relativistic objects, and galaxies'').

\bibliographystyle{apj}
\bibliography{/Users/maxim/Home/Research/BibTex,references}

\end{document}